\documentclass{pj}
\usepackage{graphics}
\usepackage{graphicx}
\usepackage{amsmath}
\usepackage{amssymb}
\usepackage{amsthm}
\usepackage{epsfig}
\usepackage{dsfont}
\begin{document}

\setcounter{page}{1}
\pjheader{Vol.\ x, y--z, 2018}

\title[Running head]
{Revisiting low-frequency susceptibility data in superconducting materials}
\footnote{\it Received 06 october 2018}  \footnote{\hskip-0.12in*\, Corresponding
author:~Jacob~Szeftel (jszeftel@lpqm.ens-cachan.fr).}
\footnote{\hskip-0.12in\textsuperscript{1} ENS Cachan, LPQM, 61 avenue du Pr\'esident Wilson, 94230 Cachan, France.\textsuperscript{2}American University of Technology, AUT Halat, Highway, Lebanon.\textsuperscript{3} Aix Marseille Univ, CNRS, Centrale Marseille, Institut Fresnel, F-13013 Marseille, France.}

\author{Jacob~Szeftel\textsuperscript{*, 1} and Michel Abou Ghantous\textsuperscript{2} and Nicolas~Sandeau\textsuperscript{3}}

\runningauthor{Szeftel and Abou Ghantous and Sandeau}

\begin{abstract}
Old  susceptibility data, measured in superconducting materials at low-frequency, are shown to be accounted for consistently  within the framework of a recently published\cite{sz1} analysis of the skin effect. Their main merit is to emphasize the  significance of  the skin-depth measurements, performed \textit{just beneath} the critical temperature $T_c$, in order to disprove an assumption, which thwarted any understanding of the skin-depth data, achieved so far by conventional high-frequency methods, so that those data might, from now on, give access to the temperature dependence of the concentration of superconducting electrons.
\end{abstract}
\setlength {\abovedisplayskip} {6pt plus 3.0pt minus 4.0pt}
\setlength {\belowdisplayskip} {6pt plus 3.0pt minus 4.0pt}
		\section{introduction}
\begin{figure}
\begin{minipage}[c]{.46\linewidth}
\label{max}
\includegraphics*[height=6 cm,width=7 cm]{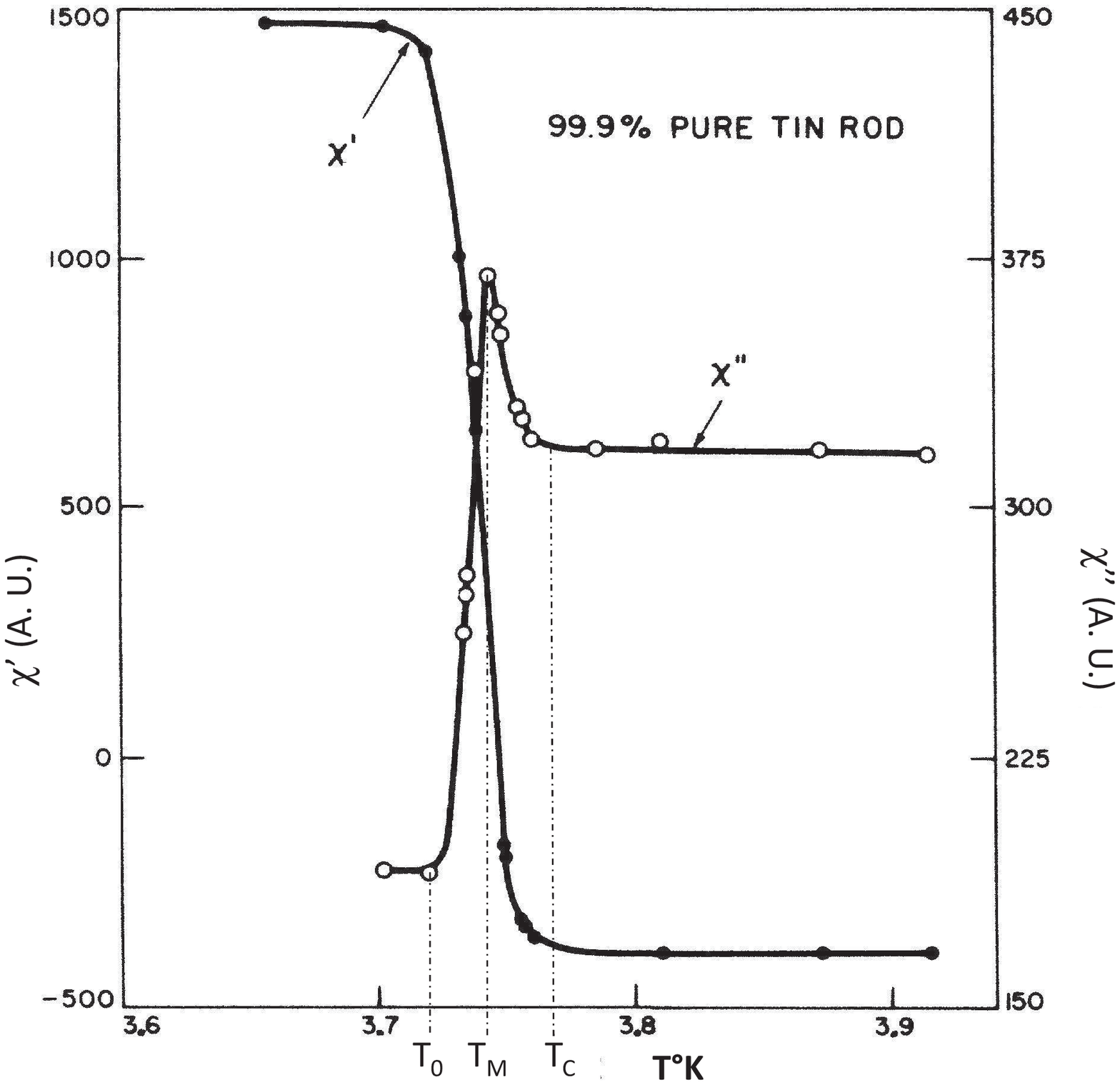}
\caption{Plot of the complex susceptibility $\chi^{'}(T)+i\chi^{''}(T)$ ($\chi^{'},\chi^{''}$ stand respectively for dispersion and absorption), as measured by Maxwell and Strongin\cite{max} in superconducting tin ($T_c=3.76K$) at $18.2$Hz; $T_M=3.74$K is the temperature associated with the maximum value of $\chi^{''}$; $T_0=3.72$K is the temperature, below which $\chi^{''}(T\leq T_0)$ remains constant}
   \end{minipage} \hfill
   \begin{minipage}[c]{.46\linewidth}
\includegraphics*[height=5 cm,width=7 cm]{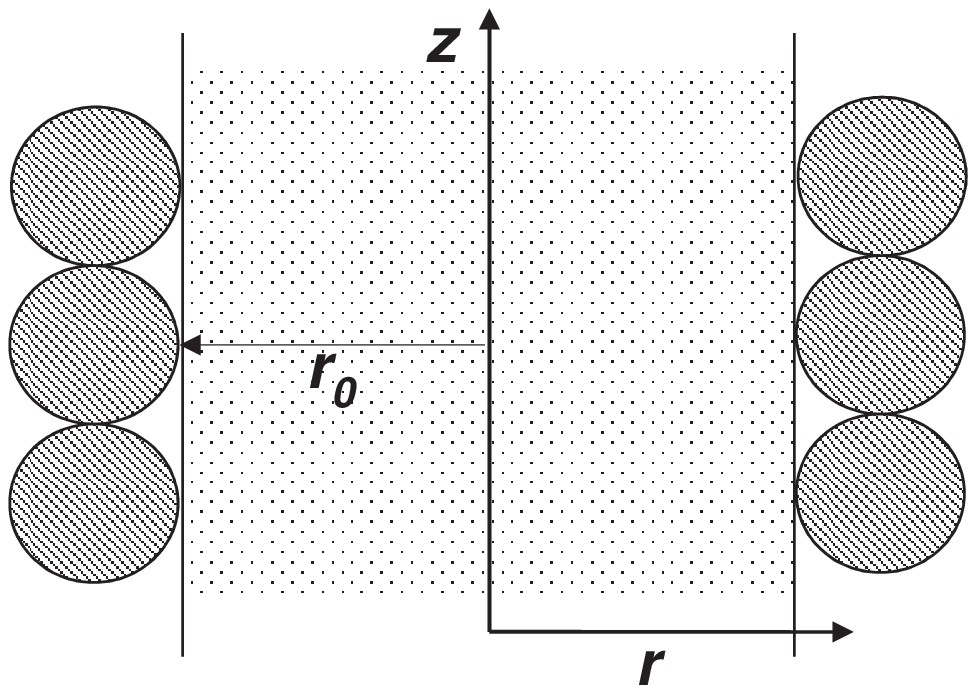}
\caption{Cross-section of the superconducting sample (dotted) and the coil (hatched); $E_\theta,j_\theta$ are both normal to the unit vectors along the $r$ and $z$ coordinates, whereas $B_z,H_z$ are parallel to the unit vector along the $z$ axis}
\label{Bzr}
\end{minipage}
\end{figure}

	Very low-frequency ($\omega<40$Hz) measurements\cite{max,str}, carried out in superconducting materials , exhibited the absorption part of the complex susceptibility rising to a maximum $\chi^{''}\left(T_M\right)$, located close to  $T_c$ (see Fig.\ref{max}), which the authors then ascribed to the vortex lattice, typical of superconductors of type II. However, as this feature was subsequently observed also in materials of type I, an alternative explanation\cite{kho}, relying heavily on the BCS gap\cite{bcs,sch}, was proposed. Anyhow, both interpretations\cite{max,str,kho} turn out to be at best qualitative and partial, since the maximum of the absorption has also been observed in gapless superconductors and even in compounds of type II in a magnetic field $H<H_{c_1}$, which warrants the absence of vortex. Moreover further low-frequency susceptibility measurements, carried out in high-$T_c$ compounds\cite{ges,sar}, were interpreted solely with help of the skin effect theory. Likewise, we shall confirm below this latter explanation, by showing \textit{quantitatively} that the low frequency behavior of the susceptibility can indeed be fully understood as a straightforward by-product of the macroscopic skin effect\cite{sz1}, valid for superconductors of both kinds as well, regardless of whether they display a BCS gap or not.\par   
	In every conductor, the real part of the dielectric constant being negative for $\omega<\omega_p$, where $\omega_p\approx 10^{16}$Hz stands for the plasma frequency, causes the electromagnetic field to remain confined within a thin layer of frequency dependent thickness $\delta(\omega)$, called the skin depth\cite{jac,bor}, and located at the outer edge of the conductor. The first measurement of $\delta$  in superconductors was done, at a \textit{single} frequency $\omega\approx 10$GHz, by Pippard\cite{pip1}, who \textit{assumed} furthermore $\omega$-independent $\delta$. The current state of affairs, regarding measurements of the skin-depth in superconductors, including both low- and high-$T_c$ materials, is muddled. On one hand, some authors\cite{h3,h4}, in the wake of Pippard's work, tend to \textit{assume} $\delta(\omega)=\lambda_L,\forall \omega$ (London's length $\lambda_L$ is defined\cite{lon} as $\lambda_L=\sqrt{\frac{m}{\mu_0 c_s e^2}}$, with $\mu_0,e,m,c_s$ standing for the magnetic permeability of vacuum, the charge, effective mass and concentration of superconducting electrons, respectively). On the other hand, low-frequency susceptbility data have been discussed\cite{ges,sar} within the framework of the skin effect theory\cite{jac,bor}, which predicts the well-known behavior $\delta(\omega)\propto 1/\sqrt{\omega}$, observed in all normal metals. Finally the conjecture $\delta(\omega)=\lambda_L,\forall \omega$, which has been questioned recently\cite{sz1,sz2}, will be rebutted below, by taking advantage of the low-frequency susceptibility data\cite{max,str,ges,sar}.\par
	This analysis will be led within the two-fluid model\cite{sch,tin}, for which the conduction electrons make up a homogeneous mixture, in thermal equilibrium at temperature $T$, of normal and superconducting electrons of respective concentrations $c_n(T),c_s(T)$, constrained for $T\leq T_c$, by $c_n(T)+c_s(T)=c_0$, with $c_0$ being the total concentration of conduction electrons. Consequently, as $T$ decreases from $T_c$ down to $T=0$, $c_n(T)$ decreases from $c_n(T_c)=c_0$, while $c_s(T)$ increases from $c_s(T_c)=0$.\par
	 The outline is as follows: the electrodynamics of the skin effect, developed elsewhere\cite{sz1}, will be recalled in Section II; this will then be used to reckon the complex susceptibility $\chi^{'}(\delta)+i\chi^{''}(\delta)$ in Section III; in Section IV, the calculated $\chi^{''}(\delta)$ and the experimental data $\chi^{''}(T)$, available in Fig.\ref{max}, will be taken advantage of to achieve $\delta(T<T_c)$, but foremost, to rebut the surmise $\delta(\omega)=\lambda_L,\forall \omega$, widely used for the interpretation of skin-depth data, obtained \cite{h3,h4} at high-frequency. The conclusions are given in Section V.\par
	Before proceeding below with the discussion of the $\chi^{''}\left(T\leq T_c\right)$ data, it is worth noticing, in Fig.\ref{max}, the dispersion $\chi^{'}\left(T\right)$ swinging abruptly at $T_c$ from $\chi^{'}\left(T>T_c\right)<0$ in the normal phase to $\chi^{'}\left(T<T_c\right)>0$ in the superconducting one. This property, which has been ascribed\cite{sz2} to the normal and superconducting states, being paramagnetic and diamagnetic, respectively, has furthermore been argued to be responsible for the Meissner effect, observed in a field-cooled sample. Therefore the gratifying agreement with the experimental evidence, displayed in Fig.\ref{max}, should contribute to ascertaining the validity of our analysis\cite{sz2} of the Meissner effect.
	\section{skin effect}
	A paramount conclusion of the study of the skin effect\cite{sz1} is that there is no difference in that respect between normal and superconducting metals. Consequently, within the two-fluid model, all of the electrodynamical properties of any superconducting material depend only on its conductivity\cite{ash} $\sigma=\frac{c_0 e^2\tau}{m}$, with $\tau$ being the decay time of the current, due to its friction on the atomic lattice. $\sigma$ is calculated as some average\cite{sz1} over the normal and superconducting conductivities $\sigma_n=\frac{c_n e^2\tau_n}{m},\sigma_s=\frac{c_s e^2\tau_s}{m}$, respectively, to be discussed later in Section IV. Besides, the superconducting decay time $\tau_s$ is \textit{finite}, as recalled by Schrieffer\cite{sch} (see\cite{sch} p.4, $2^{nd}$ paragraph, lines $9,10$: \textit{at finite temperature, there is a finite ac resistivity for all frequencies $>0$}). Moreover this property of $\tau_s$ being finite will be demonstrated in section IV.\par

	Consider as in Fig.\ref{Bzr} a superconducting material of cylindrical shape, characterized by its symmetry axis $z$ and radius $r_0$ in a cylindrical frame with coordinates ($r,\theta,z$), which has been inserted into a coil of same radius $r_0$. An oscillating current $I(t)=I_0e^{i\omega t}$, with $t$ referring to time, is fed into the coil. Then $I(t)$ induces\cite{sz1}, throughout the sample, i.e. for $r\leq r_0$, a magnetic field $H(t,r)=H_z(r)e^{i\omega t}$, parallel to the $z$ axis, and an electric field $E(t,r)=E_\theta(r)e^{i\omega t}$, normal to the unit vectors along the $r$ and $z$ coordinates. $E$ in turn induces, inside the sample, a current $j(t,r)=j_\theta(r)e^{i\omega t}$, parallel to $E_\theta$, as given by Newton's law\cite{sz1}
\begin{equation}
\label{newt}
\frac{dj}{dt}=\frac{\sigma}{\tau}E-\frac{j}{\tau}\quad,
\end{equation}
where $\frac{\sigma}{\tau}E$ and $-\frac{j}{\tau}$ are respectively proportional to the driving force accelerating the conduction electrons and a friction term, responsible for Ohm's law\cite{sz1}.\par
	The electric field $E$ and the magnetic induction $B(t,r)=\mu_0H(t,r)=B_z(r)e^{i\omega t}$, parallel to the $z$ axis (the relationship between $H,B$ reads in general $B=\mu_0\left(1+\chi\right)H$, which reduces here to $B=\mu_0H$, because of $\left|\chi\right|<<1$, as proved elsewhere\cite{sz2}) are related\cite{sz1} through the Faraday-Maxwell equation as
\begin{equation}
\label{Bz}
-\frac{\partial B}{\partial t}=\frac{E}{r}+\frac{\partial E}{\partial r}\quad.
\end{equation}
Finally the magnetic field $H$ and the current $j$ are related\cite{sz1} through the Amp\`ere-Maxwell equation as
\begin{equation}
\label{Hz}
-\frac{\partial H}{\partial r}=2j+\epsilon_0\frac{\partial E}{\partial t} \quad,
\end{equation}
with $\epsilon_0$ referring to the electric permittivity of vacuum.\par
 Replacing $E(t,r),\quad j(t,r),\quad B(t,r),\quad H(t,r)$ in Eqs.(\ref{newt},\ref{Bz},\ref{Hz}) by their time-Fourier transforms $E_\theta(\omega,r),\quad j_\theta(\omega,r),\quad B_z(\omega,r),\quad H_z(\omega,r)$ yields\cite{sz1} 
\begin{equation}
\label{fou2}
\begin{array}{l}
E_\theta\left(\omega,r\right)=\frac{1+i\omega\tau}{\sigma}j_\theta\left(\omega,r\right)\\
i\omega B_z\left(\omega,r\right)=-\left(\frac{E_\theta\left(\omega,r\right)}{r}+\frac{\partial E_\theta\left(\omega,r\right)}{\partial r}\right)\\
\frac{\partial B_z\left(\omega,r\right)}{\partial r}=-\mu_0\left(2j_\theta\left(\omega,r\right)+i\omega\epsilon_0E_\theta\left(\omega,r\right)\right)
\end{array}
\quad .
\end{equation}
Eliminating $E_\theta\left(\omega,r\right),j_\theta\left(\omega,r\right)$ from Eqs.(\ref{fou2}) gives\cite{sz1} finally
\begin{equation}
	\label{skin}
\frac{\partial^2 B_z\left(\omega,r\right)}{\partial r^2}=\frac{B_z\left(\omega,r\right)}{\delta^2(\omega)}-\frac{\partial B_z\left(\omega,r\right)}{r\partial r}\quad,\quad \delta(\omega)=\frac{\lambda_L}{\sqrt{\frac{2i\omega\tau}{1+i\omega\tau}-\frac{\omega^2}{\omega^2_p}}}\quad,
\end{equation}
with the plasma frequency defined\cite{jac,bor,tin} as $\omega_p=\sqrt{\frac{c_0 e^2}{\epsilon_0m}}$. Eqs.(\ref{skin}) yield\cite{sz1} indeed both the usual expression $|\delta|=\frac{1}{\sqrt{2\mu_0\sigma\omega}}$, valid in the low frequency limit $\omega\tau<<1$, and its lower bound $\delta=\frac{\lambda_L}{\sqrt{2}}$, reached at high frequency such that $\omega\tau>>1$.
	\section{low-frequency susceptibility}
	Due to $\delta=\frac{1}{\sqrt{2i\mu_0\sigma\omega}}$, if $\omega\tau<<1$, as inferred from Eqs.(\ref{skin}), the complex induction reads $B_z\left(u\right)=B_r\left(u\right)+iB_i\left(u\right)$, with $u=\frac{r}{\left|\delta\right|}$, so that Eq.(\ref{skin}) can be recast into 
\begin{equation}
\label{bes}
\frac{\partial^2B_r}{\partial u^2}=-B_i-\frac{\partial B_r}{u\partial u}\quad,\quad
\frac{\partial^2B_i}{\partial u^2}=B_r-\frac{\partial B_i}{u\partial u}
\quad .
\end{equation}
The system of linear differential equations in Eqs.(\ref{bes}) has been integrated over $u\in \left[0,\frac{r_0}{\left|\delta\right|}\right]$ with the following boundary conditions
$$
\dfrac{dB_r}{du}\left(0\right)=\dfrac{dB_i}{du}\left(0\right)=0\quad,\quad B_r\left(\frac{r_0}{\left|\delta\right|}\right)=B_r\left(I_0\right)\quad ,\quad  B_i\left(\frac{r_0}{\left|\delta\right|}\right)=B_i\left(I_0\right) 
\quad .$$
The complex induction $B\left(I_0\right)=B_r\left(I_0\right)+iB_i\left(I_0\right)$, induced at $r_0$ by the current $I$ flowing through the coil, has been calculated elsewhere\cite{sz1} to read $\frac{B\left(I_0\right)}{I_0}\propto 2-i\epsilon_0\rho_c\omega$, with $\rho_c$ being the resistivity of the coil.\par
\begin{figure}
\centering
\label{chilon}
\includegraphics*[height=6 cm,width=14 cm]{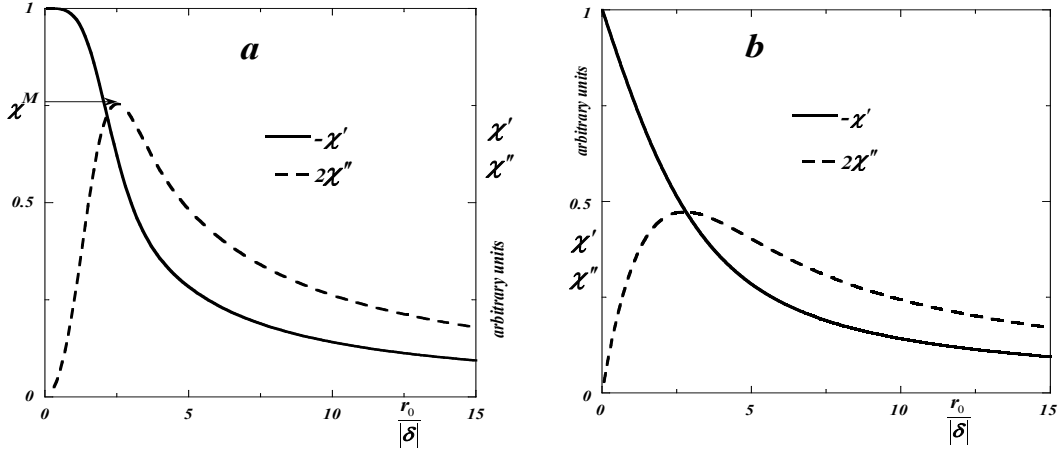}
\caption{Plots of the complex susceptibility $\chi^{'}+i\chi^{''}$ versus $\frac{r_0}{\left|\delta\right|}$, calculated with help of the solution of Eq.(\ref{skin}) and Eq.(\ref{car}), in Fig.\ref{chilon}a and Fig.\ref{chilon}b, respectively}
\end{figure}
		The complex self-inductance $L$ of the system, made up of the coil and the superconducting sample, reads as $L\propto \chi^{'}+i\chi^{''}$. Due to the very definition of $L=\frac{2\pi}{I_0}\int_{0}^{r_0}B_z\left(r\right)rdr$, the susceptibility reads finally  
$$\chi^{'}\propto\left|\delta\right|^2\int_{0}^{\frac{r_0}{\left|\delta\right|}}B_r\left(u\right)udu,\quad
\chi^{''}\propto\left|\delta\right|^2\int_{0}^{\frac{r_0}{\left|\delta\right|}}B_i\left(u\right)udu\quad.$$
	$\chi^{'}\left(\frac{r_0}{\left|\delta\right|}\right),\chi^{''}\left(\frac{r_0}{\left|\delta\right|}\right)$ have been plotted in Fig.\ref{chilon}a. The characteristic features, mentioned elsewhere\cite{max,str,ges,sar}, can be seen conspicuously, namely $\chi^{'}\left(\frac{r_0}{\left|\delta\right|}\right)$ decreases monotonically, whereas $\chi^{''}\left(\frac{r_0}{\left|\delta\right|}\right)$ goes through a maximum $\chi^M$ at $\frac{r_0}{\left|\delta\right|}=2.53$. The maximum was reported\cite{max,str} previously to show up rather at $\frac{r_0}{\left|\delta\right|}\approx 1.8$. This discrepancy is to be ascribed to a different definition of the susceptibility, chosen there\cite{max,str,kho}, which nevertheless does not correspond, unlike the definition used here, to what is actually measured in the experimental procedure, i.e. the complex impedance of the circuit, comprising the coil and the sample.  The data in Fig.\ref{chilon}a are independent from $\omega$ because $\omega$ shows up only in the expression of $B_i\left(I_0\right)$, which turns out to be negligible ($\left|B_i\left(I_0\right)\right|\approx 10^{-18}$ for $\omega=100$Hz). However they do \textit{depend} on the sample \textit{shape}, as illustrated by reckoning the susceptibility for the semi-infinite geometry considered by London\cite{lon}. In this latter case, Eq.(\ref{skin}) should be replaced by 
\begin{equation}
\label{car}
\frac{\partial^2 B_z}{\partial r^2}=\frac{B_z}{\delta^2}\Rightarrow B_z(u)=2e^{\frac{1+i}{\sqrt{2}}\left(u-\frac{r_0}{\left|\delta\right|}\right)} \quad,
\end{equation}
which entails that $\chi^{'},\chi^{''}$ read now
$$
\chi^{'}\propto \frac{\left|\delta\right|}{\sqrt{2}r_0}-\frac{e^{-\frac{r_0}{\sqrt{2}\left|\delta\right|}\textrm{sin}\left(\frac{r_0}{\sqrt{2}\left|\delta\right|}\right)}}{\left(\frac{r_0}{\sqrt{2}\left|\delta\right|}\right)^2} \quad,\quad
\chi^{''}\propto -\frac{\left|\delta\right|}{\sqrt{2}r_0}+\frac{1-e^{-\frac{r_0}{\sqrt{2}\left|\delta\right|}\textrm{cos}\left(\frac{r_0}{\sqrt{2}\left|\delta\right|}\right)}}{\left(\frac{r_0}{\sqrt{2}\left|\delta\right|}\right)^2}
\quad .$$
The corresponding $\chi^{'}\left(\frac{r_0}{\left|\delta\right|}\right),\chi^{''}\left(\frac{r_0}{\left|\delta\right|}\right)$ data are depicted in Fig.\ref{chilon}b. Note that the maximum value of $\chi^{''}$ shifts from $\frac{r_0}{\left|\delta\right|}=2.53$ in Fig.\ref{chilon}a up to $\frac{r_0}{\left|\delta\right|}=2.77$ in Fig.\ref{chilon}b. Likewise the ratios $\chi^{'}/\chi^{''}$ differ from each other for any $r_0/|\delta|$ in both figures. Those differences stem from the respective solutions $B_z\left(u\right)$ of Eqs.(\ref{skin},\ref{car}) deviating from each other in the relevant domain, i.e. for $u\approx 1$, even though they are practically identical for $u >>1$, because the solution of Eq.(\ref{skin}) is a Bessel function, having the property $B_z(r)\approx e^{r/\delta(\omega)}$ if $r>>|\delta(\omega)|$.
	\section{experimental discussion}
\begin{figure}
\label{delta}
\centering
\includegraphics*[height=6 cm,width=6 cm]{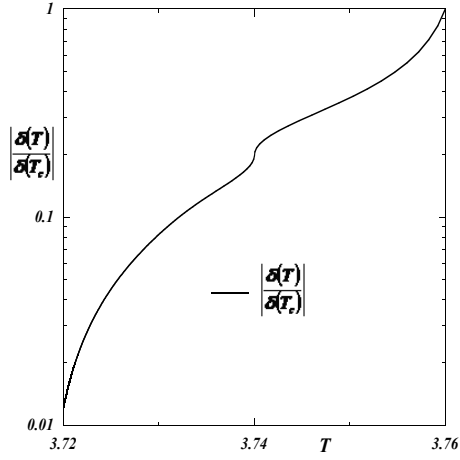}
\caption{Semilogarithmic plot of $\left|\delta\left(T\in\left[T_0,T_c\right]\right)\right|$ with $T_0=3.72$K and $T_c=3.76$K; the discontinuity of $\frac{d\left|\delta\right|}{dT}$ at $T_M=3.74$K, which reflects that of $\frac{d\chi^{''}}{dT}\left(T_M\right)$ in Fig.\ref{max}, is a spurious effect, caused by the experimental difficulty\cite{max} in resolving the narrow temperature range $\left[T_0,T_c\right]$} 
\end{figure}

	We shall now take advantage of both $\chi^{''}\left(T\right)$ data, taken from Fig.\ref{max}, and  $\chi^{''}\left(\delta\right)$ ones, pictured in Fig.\ref{chilon}a, to chart $\delta\left(T\leq T_c\right)$. The one to one correspondence between $\chi^{''}\left(T\right),\chi^{''}\left(\delta\right)$ is then expressed as
$$
\frac{\chi^{''}\left(T>T_M\right)-\chi^{''}\left(T_M\right)}{\chi^{''}\left(T_c\right)-\chi^{''}\left(T_M\right)}=\frac{\chi^{''}\left(\delta\left(T\right)\right)-\chi^M}
{\chi^{''}\left(\delta\left(T_c\right)\right)-\chi^M}\quad,\quad
\frac{\chi^{''}\left(T<T_M\right)-\chi^{''}\left(T_M\right)}{\chi^{''}\left(T_0\right)-\chi^{''}\left(T_M\right)}=\frac{\chi^{''}\left(\delta\left(T\right)\right)-\chi^M}
{\chi^{''}\left(\delta\left(T_0\right)\right)-\chi^M}
\quad .$$
$T_M,T_0$ are defined in the caption of Fig.\ref{max} and $\chi^M$ refers to the maximum value of  $\chi^{''}\left(\delta\right)$ (see Fig.\ref{chilon}a). As the values of $\delta\left(T_c\right),\delta\left(T_0\right)$ are unknown, we have assumed arbitrarily $\frac{r_0}{\left|\delta\left(T_c\right)\right|}=.5,\frac{r_0}{\left|\delta\left(T_0\right)\right|}=41$, in order to proceed with the detailed discussion of an illustrative example. Finally the resulting $\left|\delta\left(T\leq T_c\right)\right|$ curve has been plotted in Fig.\ref{delta}. Its prominent feature is the large variation of $|\delta|$, and thence of $\sigma$, over a narrow temperature range $T_c-T_0=.04$K, resulting from the steep decrease of $c_s$ down to $0$ for $T\rightarrow T_c^-$, as shown below.\par
	 In the two-fluid model, $\sigma$ reads\cite{sz1}, at low frequency such that $\omega\tau_s<<1$, as
\begin{equation}
\label{sig}
\sigma\left(T\right)=\sigma_n+\sigma_s=\frac{e^2}{m}\left(\left(c_0-c_s\left(T\right)\right)\tau_n+c_s\left(T\right)\tau_s\right)\quad ,
\end{equation}
because of $c_n\left(T\leq T_c\right)+c_s\left(T\right)=c_0$. Consequently, the average $\tau$, showing up in Eq.(\ref{newt}), is defined as $c_0\tau=\left(c_0-c_s\right)\tau_n+c_s\tau_s$. At low $T$, the current decay is due to	scattering by impurities and (or) dislocations, so that $\tau_n,\tau_s$ are $T$-independent. Although the value of $\tau$ is in general unknown, it could be measured as indicated elsewhere\cite{sz1}. Then the measurement of $\delta$ would provide with a rather \textit{unique} access to $c_s\left(T\right)$. The steep increase of $\left|\delta\right|$ for $T\rightarrow T_c^-$, seen in Fig.\ref{delta}, stems from the rapid decrease of $c_s\left(T\rightarrow T_c^-\right)\rightarrow 0$ \textit{with} $\tau_n<<\tau_s$\cite{sz1}, consistently with Eq.(\ref{sig}).\par
	  Besides, both the observed\cite{str,ges,sar} downward shift of $T_M\searrow 0$ with growing magnetic field $H$ or impurity concentration and, conversely, the upward one $T_M\nearrow T_c$, resulting from increasing the frequency $\omega$, are very well accounted for within the framework of this analysis :
\begin{itemize}
	\item
	increasing $H$ has been shown to cause $c_s$ (see arXiv : 1704.03729), and thence $\sigma$ (see Eq.(\ref{sig})) to decrease. Due to $|\delta|=\frac{1}{\sqrt{2\mu_0\sigma\omega}}$, decreasing $\sigma$ leads in turn to increasing $|\delta|$. Because the maximum of $\chi^{''}\left(\delta\right)$ is pinned at $\frac{r_0}{\left|\delta\right|}=2.53$ (see Fig.\ref{chilon}a), the maximum of $\chi^{''}\left(T_M=T_i(H=0)\right)$ will be shifted down to $T_M=T_f(H\neq 0)<T_i$, owing to $c_s(T_f<T_i)>c_s(T_i)\Rightarrow \sigma(T_f)>\sigma(T_i)$, such that $\left|\delta(H=0,T_i)\right|=\left|\delta(H\neq 0,T_f)\right|\Rightarrow\frac{r_0}{\left|\delta(H=0,T_i)\right|}=\frac{r_0}{\left|\delta(H\neq 0,T_f)\right|}=2.53$. Likewise, as increasing the impurity concentration causes both $\tau$ and thence $\sigma$ to decrease, the same rationale implies that the maximum of $\chi^{''}(T)$ will be pushed towards lower $T$ too; 	
	\item 
	conversely, because of $\left|\delta(\omega)\right|\propto 1/\sqrt{\omega}$, growing $\omega$ from $\omega_i$ up to $\omega_f>\omega_i$ will shift the temperature $T_i$ of the maximum of $\chi^{''}\left(T_M=T_i(\omega_i)\right)$ up to $T_f(\omega_f>\omega_i)>T_i$, so that $\frac{r_0}{\left|\delta(\omega_i,T_i)\right|}=\frac{r_0}{\left|\delta(\omega_f,T_f)\right|}=2.53$.
\end{itemize}\par
	 For low-frequency measurements of $\chi^{''}\left(T\right)$ to be useful, the prerequisite  $\left|\delta(\omega,T_c)\right|> r_0$ must be fulfilled. Such a condition requires, in very pure superconductors ($\Rightarrow$ large $\tau$-value), to operate at unpractical low frequency. This is the real reason why Maxwell and Strongin\cite{max} failed to observe any maximum of $\chi^{''}\left(T\right)$ in very pure $Sn$. This drawback and the additional one that the values of $\delta\left(\omega,T_c\right),\delta\left(\omega,T_0\right)$ are in general unknown, entail that the measurement of the low-frequency susceptibility could not be regarded as a practical alternative method to conventional high-frequency ones\cite{h3,h4}, currently used to measure the skin-depth. However, due to a special circumstance to be discussed now, it still turns out to be of great relevance.\par
	At high enough a frequency, such that $r_0>>\left|\delta\left(\omega\right)\right|$, it has been shown\cite{sz1} that $\left|\delta\right|=C(\omega)\chi^{'}(\omega,T)$, for which $C(\omega)$ is an unknown experimental coefficient. Therefore, assessing the accurate value of $\delta(T)$ requires to perform the measurement of $\chi^{'}(T)$ up to $T_c$, because the value of $\sigma(T_c)=\sigma_n(T_c)$ and thence that of $|\delta(T_c)|=\frac{1}{\sqrt{2\mu_0\sigma_n(T_c)\omega}}$ are well known. However, an accurate assignment of $\delta\left(\omega\right)$ is still lacking, because all published $\delta\left(\omega\right)$ data have so far been obtained from $T=0$ up to $T_0<T_c$, by consistently refraining from publishing $\chi^{'}(T\in\left[T_0,T_c\right])$, whereas $\chi^{''}(T\in\left[T_0,T_c\right])$ data, albeit of little usefulness, are conversely \textit{available} in the literature. This harmful state of affairs results from a long standing contradiction between the assumption $\delta\left(\omega\right)=\lambda_L,\forall \omega$, used in the interpretation of all experimental $\delta(T<T_c)$-data and the experimental evidence itself. It is thus in order to resolve this contradiction.\par
	 The expression of the average $\delta$ in the two-fluid model has been shown\cite{sz1} to read 
\begin{equation}
\label{skin3}
\delta^{-2}=2i\mu_0\omega\left(\frac{\sigma_n}{1+i\omega\tau_n}+\frac{\sigma_s}{1+i\omega\tau_s}\right)\quad.
\end{equation}
As a consequence of the mainstream assumption\cite{tin} $\tau_s\rightarrow\infty$, Eqs.(\ref{newt},\ref{skin3}) lead indeed to the complex impedance of the sample, reading as $Z=\sigma_n\left(T\right)-i\omega\mu_0\lambda_L^2$, and to $\delta\left(\omega\right)=\lambda_L/\sqrt{2},\forall \omega$, respectively. Thence the average conductivity $\sigma$, equal to the real part of $Z$, is inferred to read, at low frequency such that $\omega\tau_n<<1$, as  $\sigma\left(T<T_c\right)=\sigma_n\left(T\right)=\frac{c_n\left(T\right)e^2\tau_n}{m}\Rightarrow\sigma\left(T<T_c\right)<\sigma_n\left(T_c\right)=\frac{c_0e^2\tau_n}{m}$, because of $c_0>c_n\left(T<T_c\right)=c_0-c_s\left(T\right)$. As a matter of fact, both mainstream claims $\delta\left(\omega\right)=\frac{\lambda_L}{\sqrt{2}},\forall\omega$ and $\sigma\left(T<T_c\right)<\sigma_n\left(T_c\right)$ are seen to run afoul at experimental evidence :
\begin{itemize}
	\item
	the data, pictured in Fig.\ref{chilon}a, show that $\delta\left(\omega=18Hz\right)\approx r_0>10^{-4}m$, whereas $\lambda_L=\sqrt{\frac{m}{\mu_0 c_s e^2}}<10^{-7}m$, so that $\delta\left(\omega=18Hz\right)>>\lambda_L$, which disproves the mainstream claim $\delta\left(\omega\right)=\lambda_L,\forall \omega$;
	\item
	the measured, average ac conductivity for the superconducting phase of $YBa_2Cu_30_7$ has been reported\cite{ges,sar} to be $\sigma\left(T<T_c\right)\approx 10^5\sigma_n\left(T_c\right)\Rightarrow\sigma\left(T<T_c\right)>>\sigma_n\left(T_c\right)$, which rebuts the opposite claim $\sigma\left(T<T_c\right)<\sigma_n\left(T_c\right)$, inferred above from the mainstream view. 
\end{itemize}\par
	In summary, the only way to extract useful information from the $\chi^{'}$-data, obtained at high frequency $\omega\in\left[10MHz,10GHz\right]$, is to measure $\chi^{'}(T)$ \textit{up to} $T=T_c$ at \textit{two} frequencies, \textit{distant} from each other, as advised elsewhere\cite{sz1}.\\
	\section{conclusion}
	The measured low-frequency susceptibility data in superconducting materials have been comprehensively explained within a recent account of the skin effect\cite{sz1}. The prominent maximum of the absorption $\chi^{''}(T)$ has been associated with the steep decrease of the concentration of superconducting electrons $c_s(T)\rightarrow 0$ for $T\rightarrow T_c^-$ \textit{with} the prerequisite $\left|\delta(\omega,T_c)\right|> r_0$, whereas the dispersion $\chi^{'}$ changing its sign  at $T_c$ has been identified to be the driving force of the Meissner effect, observed in a field-cooled sample\cite{sz2}.\par
	 Two remarks are of interest :
\begin{itemize}
	\item
	by contrast with a normal metal, for which the temperature behavior of $\sigma=\frac{c_0 e^2\tau_n}{m}$ is completely determined by $\tau_n(T)$, as $c_0$ remains constant, $\sigma(T<T_c)$ in a superconductor is closely related to the variation of $c_s(T)$, because $\tau_s$ is almost $T$-independent;
	\item
	 the high frequency $\delta$ assignments are obtained through the measurement of the \textit{dispersion} $\chi^{'}$, while the low frequency ones are obtained through that of the \textit{absorption} $\chi^{''}$.
\end{itemize}
Finally, the assumption $\delta\left(\omega\right)=\lambda_L,\quad\forall\omega$ has been rebutted, which would enable one to monitor, thanks to high frequency determination of $\delta$, $c_s(T<T_c)$ and also the $c_s$ dependence on the applied magnetic field or the induced persistent current during the reversible superconducting-normal transition (see arXiv : 1704.03729).  

\end{document}